\documentclass[a4paper,12pt]{article}
\usepackage{amssymb,latexsym,amsmath}

\usepackage{fullpage}
\usepackage{mathrsfs}
\usepackage{bbold}

\usepackage[final]{hyperref}
\hypersetup{pdfborder={0 0 0},colorlinks=false,urlcolor=red!60!black,citecolor=blue!60!black,linkcolor=red!60!black,linktocpage=true}

\usepackage[pdftex]{graphicx}
\usepackage[table,dvipsnames]{xcolor}
\usepackage{multirow}
\usepackage{rotating}
\usepackage{array}
\usepackage{float}
\restylefloat{table}

\makeatletter \@addtoreset{equation}{section}
\makeatother

\begin{document}

\begin{titlepage}
	\thispagestyle{empty}
	\begin{flushright}
		\hfill{QMUL-PH-20-34}
	\end{flushright}
		
	\begin{center}

		{ \LARGE{\bf Old and new vacua \\[4mm]
		of 5D maximal supergravity}}
	
		\vspace{50pt}
							
			{G.~Dall'Agata$^{1,2}$, G.~Inverso$^{2,3}$ and D.~Partipilo$^{1,2}$}
							
			\vspace{25pt}
							
			{
				{\it  $^{1}$Dipartimento di Fisica e Astronomia ``Galileo Galilei''\\
					Universit\`a di Padova, Via Marzolo 8, 35131 Padova, Italy}
										
				\vspace{15pt}
										
				{\it   $^{2}$INFN, Sezione di Padova \\
				Via Marzolo 8, 35131 Padova, Italy}

				\vspace{15pt}

				{\it $^3$Centre for Research in String Theory,\\
				School of Physics and Astronomy, Queen Mary University of London,\\
				327 Mile End Road, London, E1 4NS, United Kingdom}
			}

		\vspace{80pt}

		{ABSTRACT}
	\end{center}
	
	\vspace{10pt}
	
	We look for critical points with U(2) residual symmetry in 5-dimensional maximally supersymmetric gauged supergravity, by varying the embedding tensor, rather than directly minimizing the scalar potential. 
	We recover all previously known vacua and we find four new vacua, with different gauge groups and cosmological constants. 
	We provide the first example of a maximal supergravity model in $D \ge 4$ having critical points with both positive and vanishing cosmological constant.
	For each vacuum we also compute the full mass spectrum. All results are analytic.

\end{titlepage}

\baselineskip 6 mm

\allowdisplaybreaks

\date{}

\section{Introduction} 
\label{sec:introduction}

Charting and analyzing vacua of supergravity theories is a fundamental task to find which models can be related to string theory as well as to understand supersymmetry breaking, the possible mechanisms to generate critical points with a positive value of the cosmological constant and which supergravities lead to Anti-de Sitter (AdS) vacua with an interesting holographic dual.
Among all possible theories, the maximally supersymmetric ones stand out for their fixed matter content and the limited number of possible deformations.
For these reasons there has been an active interest in their gaugings and in the analysis of the resulting scalar potentials to understand their critical points, with a special emphasis on the theories obtained by reducing string or M-theory on spheres, which give models with vacua dual to maximally supersymmetric Conformal Field Theories (CFT).

The main challenges one faces when dealing with this problem are associated to the very complicated structure of the scalar potential, a function of 70 or 42 scalars in the maximal theory in 4 and 5  dimensions respectively, which also depends on a large number of parameters (912 and 351 respectively) that fix the structure of the gaugings and therefore of the full lagrangian, according to the rules specified in \cite{deWit:2004nw,deWit:2007kvg}.
Clearly such a large space of parameters makes the search for critical points complicated and attempts at a general classification extremely difficult.
However, there has been some interesting progress in the last few years that expanded a lot our knowledge of this particular aspect of maximal supergravity theories.

There are mainly three techniques that have been used so far to find and analyze critical points of (maximal) supergravity theories.
The first one relies on using symmetries to consistently truncate a particular theory to a subset of fields containing a limited number of scalars and then extremising the resulting simplified potential.
Pioneered in \cite{Warner:1983du,Warner:1983vz}, this technique allowed for the first and only analytic results for the maximal theories from the '80s until recent years.
For what concerns maximal supergravity in 5 dimensions, this technique allowed the discovery of 5 vacua \cite{Gunaydin:1985cu, Girardello:1998pd,Distler:1998gb,Khavaev:1998fb} of the SO(6) and SO(3,3) gauged models in addition to the maximally symmetric one in \cite{Pernici:1985ju}, though often only partial results were available on the spectrum about these vacua.

More recently, a new numerical approach, based on Machine Learning software libraries was developed and employed in a series of papers \cite{Fischbacher:2009cj,Fischbacher:2010ki,Fischbacher:2011jx,Comsa:2019rcz,Bobev:2019dik,Krishnan:2020sfg,Bobev:2020ttg,Bobev:2020qev} where many new vacua of the maximal supergravities in 4 and 5 dimensions had been found.
This also allowed to find precise information about the spectrum of scalar fluctuations, residual gauge groups and residual supersymmetry.
In particular, 27 new AdS vacua were found in the SO(6) maximal supergravity in 5 dimensions, with a detailed analysis in \cite{Krishnan:2020sfg,Bobev:2020ttg}.

While these approaches are very interesting and gave promising results, so far they have only been used to produce critical points for a fixed scalar potential, which is resulting from a single specific gauging within the large infinite family of possible deformations.
This leaves open the possibility that other vacua with the same residual symmetries appear in different gaugings.
The approach we are going to use in this work uses instead the power of the embedding tensor formalism in a way that allows for the search of critical points independently from the choice of gauging.
This approach was pioneered in a very different context in \cite{Li:1986tk} and used in the context of maximal 4-dimensional supergravity in \cite{DallAgata:2011aa,DallAgata:2012mfj,Borghese:2012qm,Borghese:2012zs,Borghese:2013dja,Catino:2013ppa,Gallerati:2014xra}, 
as well as in half-maximal supergravity in four and three dimensions \cite{Dibitetto:2011gm,Deger:2019tem}.
In addition to the power of investigating in a single sweep all deformations of maximal supergravity, this approach has so far produced analytic results for the critical points and their full spectrum, also providing information on the gauging, the residual gauge symmetry and supersymmetry of the vacua.
Moreover, for Minkowski vacua this led to understanding the moduli space of these theories \cite{Catino:2013ppa} as well as their uplift to string theory \cite{Inverso:2019mnv}.
Finally, since the vacua are obtained without specifying first the gauging, this means that we can exhaustively classify vacua with a given residual symmetry for all possible consistent gaugings.

In this work we apply this last technique by investigating critical points of maximal gauged supergravities in 5 dimensions with a residual U(2) symmetry.
We recover all previously known vacua and we find four new ones, with different gauge groups and cosmological constants.
We also provide analytic results for their full mass spectra, thus completing partial results for old vacua as well as fully analyzing new ones.
We did not find new AdS vacua, so that the only such vacua with U(2) symmetry are those appearing in the maximal supergravity with SO(6) gauge group, but we have new Minkowski and de Sitter vacua.
A particularly interesting result is that two of the vacua appear in the same theory with SO$^*(6) = $SU(3,1) gauge group, providing the first example n $D \ge 4$ where a single gauging of a maximal supergravity theory produces vacua in different classes of the cosmological constant, one having a positive cosmological constant and the other a vanishing cosmological constant and residual supersymmetry\footnote{There is a similar instance in maximal supergravity in 3 dimensions \cite{Fischbacher:2002fx}}.

In what follows, after some summary of the main ingredients of maximal supergravity in 5 dimensions, we will discuss in some detail our technique in section \ref{sec:extrema_of_the_scalar_potential} and then proceed with a detailed analysis of the U(2) invariant sector in section \ref{sec:vacua_with_residual_u_2_symmetry}.
We tried to summarize all our results in tables that could be easily consulted and used for future reference.


\section{N=8 supergravity in 5 dimensions} 
\label{sec:n_8_supergravity_in_5_dimensions}

A comprehensive presentation of the 5-dimensional $N=8$ supergravity lagrangian, supersymmetry rules and of all the details regarding the gauging procedure can be found in \cite{deWit:2004nw}, which we are going to use as a basis for our analysis.
In order to facilitate reading, we collected in this section the main formulas and properties of the tensors relevant for our work.

Any gauging of 5-dimensional maximal supergravity is specified by the choice of the embedding tensor $\Theta_M{}^\alpha$, which selects the generators $t_{\alpha}$ of the duality algebra $\mathfrak{e}_{6(6)}$  associated to the vector fields $A_\mu^M$ gauging the corresponding group.
Once this tensor is fixed, everything else in the lagrangian and supersymmetry transformations follows, according to the analysis in \cite{deWit:2004nw}.
The embedding tensor lives in the product representation $\mathbf{27} \times \mathbf{78}$ of E$_{6(6)}$, but it is constrained by supersymmetry and gauge-invariance to the representation $\mathbf{351}$, which is then further constrained by consistency conditions quadratic in the embedding tensor.

These \textbf{351} parameters can also be codified in a different set of tensors that explicitly transform under the maximal compact subgroup of E$_{6(6)}$, namely USp(8).
Since we are interested in the critical points of the scalar potential, we will in fact make extensive use of the fermion shifts
\begin{equation}
	A_{ij} = A_{ji} = \Omega_{ik} \Omega_{jl} A^{kl} = (A^{ij})^*
\end{equation}
and
\begin{equation}
	A_{i,jkl} = A_{i,[jkl]} = \Omega_{im} \Omega_{jn} \Omega_{kp} \Omega_{lq} A^{m,npq} = (A^{i,jkl})^*,
\end{equation}
satisfying
\begin{equation}
	A_{i,jkl} \Omega^{kl} = 0 = A_{[i,jkl]} = \Omega^{ij} A_{i,jkl}.
\end{equation}
These tensors are expressed in terms of the USp(8) indices $i,j,\ldots = 1,\ldots,8$, which are carried by the fermion fields and by the supersymmetry parameters $\epsilon^i$ ($\Omega$ is the USp(8)  symplectic-invariant form).
In fact $A_{ij}$ is in the representation $\mathbf{36}$ and $A_{i,jkl}$ in the $\mathbf{315}$ of USp(8), which are precisely the representations under which the $\mathbf{351}$ of E$_{6(6)}$ breaks.  
They are called fermion shifts because they appear in the supersymmetry transformations of the fermion fields and are non-vanishing only when there is a non-trivial gauging, hence shifting the ungauged expression.

The relation between the fermion shifts and the embedding tensor is expressed via the T-tensors:
\begin{eqnarray}
T^{klmn}{}_{ij} &=& 4\, A_2{}^{q,[klm}\, \delta^{n]}{}_{[i}\, \Omega_{j]q} + 3\, A_2{}^{p,q[kl}\,\Omega^{mn]}\, \Omega_{p[i}\,\Omega_{j]q}  \,, \\[2mm]
T_i{}^{jkl} &=& -\Omega_{im}\,A_2{}^{(m,j)kl} - \Omega_{im} \Big(\Omega^{m[k} \,A_1{}^{l]j} +\Omega^{j[k} \,A_1{}^{l]m}+ \frac14 \Omega^{kl}\,A_1{}^{mj} \Big) \,, \\[2mm]
  {\cal Z}^{ij,kl} &=& \Omega^{[i[k}\,A_1{}^{l]j]} + A_2{}^{[i,j]kl} \,,
\end{eqnarray}
and can be written as
\begin{equation}
		X_{MN}{}^P = \Theta_{M}{}^\alpha (t_{\alpha})_N{}^P = {\cal V}_M^{mn} {\cal V}_N{}^{kl} {\cal V}_{ij}^P\,\left[2\, \delta_k^i\, T^j{}_{lmn} +T^{ijpq}{}_{mn}\, \Omega_{pk}\, \Omega_{ql}\right],
\end{equation}
where ${\cal V}_M^{ij}$ are the coset representatives of the E$_{6(6)}/$USp(8) scalar manifold, satisfying ${\cal V}_M^{ij} \Omega_{ij} = 0$, and ${\cal V}^M_{ij}$ are their inverse ${\cal V}_M^{ij}\, {\cal V}_{ij}^N = \delta_M^N$.

Once we use the T-tensor, the quadratic constraints on the embedding tensor have a rather simple expression:
\begin{equation}\label{quadconst}
   T^i{}_{jkl}\, {\cal Z}^{kl,mn} =0= T^{ijkl}{}_{mn}\,{\cal Z}^{mn,pq}\,.
\end{equation}

Let us now come to the center of our analysis: the scalar potential and the mass matrices.
While everything can be defined in terms of the fermion shifts, for the scalar masses we preferred to use a convenient expression which is valid only at the selected point of the scalar manifold we use as a basis for our analysis.
As we will see we are not going to lose generality by this assumption.

Following a well-known general rule of gauged supergravity theories, the scalar potential is the square of the fermion shifts:
\begin{equation}
	V = 3 \, A^{ij} A_{ij}   - \frac13\,A^{i,jkl}A_{i,jkl}.
\end{equation}
We are looking for maximally-symmetric vacua, where all fields are vanishing except for the scalar fields, which could have a constant vacuum expectation value and for the metric, which either describes a de Sitter, Minkowski of anti-de Sitter spacetime.
The scalar equations of motion are solved by the critical point condition
\begin{equation}\label{eom}
	U_{ijkl} - \frac{3}{2} \Omega_{[ij}\, U_{kl]pq}\,\Omega^{pq} + \frac{1}{8}\,\left(U_{mnpq} \,\Omega^{mn}\, \Omega^{pq}\right)  \,\Omega_{[ij}\Omega_{kl]}= 0\,,
\end{equation}
where the tensor $U_{ijkl}$ is 
\begin{equation}
U_{ijkl}= \frac43 \,A_1{}^{mq} \,A_{2\,m,[ijk}\,\Omega_{l]q}  +2\,  A_{2}{}^{m,npq} \, A_{2\,n,m[ij}\,\Omega_{|p|k}\, \Omega_{l]q}  \,.
\end{equation}

Once we find a critical point, we derive the masses of the various fields by computing the eigenvalues of the respective mass matrices.
For what concerns the gravitini $\psi_\mu^i$, the mass matrix is directly proportional to the $A_{ij}$ shift matrix
\begin{equation}
		{\cal M}^{(3/2)}_{ij} = \frac32\, A_{ij}.
\end{equation}
The masses of the other fermions $\chi^{ijk}$ are then fixed by the eigenvalues of (indices $ijk$ and $pqr$ are fully antisymmetrized)
\begin{equation}
		{\cal M}^{(1/2)}_{ijk,pqr} = 8\, A_{[i,j[pq} \Omega_{r]k]} + 2\, A_{[i[p} \Omega_{qj} \Omega_{k]r]} -\frac{10}{3} \, A_{l,\,ijk} A^{lm} (A_{m,stu} A^{n,stu})^{-1} A_{n,pqr}.
\end{equation}
This mass matrix is the result of subtracting from the lagrangian mass the appropriate term to remove the goldstinos from the spectrum for susy-breaking vacua.
It is understood that in case of a degenerate matrix $A_{m,stu} A^{n,stu}$, we only compute the inverse for its non-degenerate part, as this is the part related to the goldstino directions, which in the original lagrangian mix the gravitinos and the spin-1/2 fields.
The proof that such additional term correctly produces ${\cal M}^{1/2}_{ijk,pqr}A^{s,pqr} = 0$ follows once one takes into account the equations of motion (\ref{eom}) and one uses repeatedly the quadratic constraints (\ref{quadconst}). 
In particular the matrix we are inverting is related to the shift of the gravitinos by means of the quadratic identity known as supersymmetric ward identity
\begin{equation}
	\frac13\, A_{j,stu} A^{i,stu} = \frac18\, \delta^i_j \,V + 3 \,A^{ip} A_{pj},
\end{equation}
which also tells us that the expression is explicitly dependent on the value of the cosmological constant at the vacuum.
This expression generalizes previous similar formulae for maximal theories in 4 dimensions, which were obtained in particular instances where the cosmological constant was vanishing \cite{DallAgata:2012tne} or when the squared shifts had already been diagonalized \cite{Gallerati:2014xra}.
A simple way to understand this expression can also be obtained by comparing it with the analogous expression for $N=1$ supergravity presented in \cite{Ferrara:2016ntj}.

Also the masses of the bosonic degrees of freedom can be expressed in terms of the same tensors.
The vector mass matrix is
\begin{equation}
	{\cal M}_{(v)M}{}^{N} = \frac13 \, {\cal V}_M^{ij} \, T^{mnpq}{}_{ij} T_{mnpq}{}^{kl} {\cal V}_{kl}^{N},
\end{equation}
while the squared masses of the tensor fields follow from the eigenvalues of the matrix
\begin{equation}
	{\cal M}_{(t)M}{}^N =  {\cal V}_M^{ij} \, {\cal Z}_{ij\,mn} {\cal Z}^{mn\,kl} {\cal V}_{kl}^{N}.
\end{equation}
These mass matrices are clearly redundant, because the sum of vector and tensor fields present in the theory is fixed, given that the tensor fields appear by dualization of the vector fields.
This means that both $M_{(v)}$ and $M_{(t)}$ are degenerate and contain zeros in the directions where the fields have been dualized.

All the above expressions have general validity and should be evaluated at the critical points satisfying (\ref{eom}).
For the scalar fields, on the other hand, following \cite{DallAgata:2011aa} we provide an expression that is valid only when the critical point is the base-point of the manifold, i.e.~when all scalars are vanishing.
While this could seem a restriction, as we will explain in the next section, it allows us to obtain the full spectrum for any critical point in any arbitrary gauging.
This is given in terms of the embedding tensor, the $\mathfrak{e}_{6(6)}$ generators, the $\mathfrak{e}_{6(6)}$ structure constants $f_{\alpha \beta}{}^\gamma$ and the $\mathfrak{e}_{6(6)}$ Cartan--Killing metric $\eta_{\alpha \beta}$:
\begin{equation}
	\begin{split}
	{\cal M}_{\alpha}{}^\beta = &\frac{16}{5}\left(\Theta_M{}^\sigma (t_{\alpha}t^\beta)_M{}^N \Theta_N{}^\gamma (\delta^\gamma_ \sigma + 5\, \eta_{\sigma \gamma}) + \Theta_M{}^\sigma (t_{\alpha})_M{}^N \Theta_N{}^\gamma f^{\beta}{}_{\gamma}{}^\sigma \right.\\[2mm]
	&\left.+ \Theta_M{}^\sigma (t^{\beta})_M{}^N \Theta_N{}^\gamma f{}_{\alpha\gamma}{}^\sigma+ \Theta_M{}^\sigma \Theta_M{}^\gamma f_{\alpha \gamma}{}^\delta f{}^\beta{}_{\delta}{}^\sigma\right)\,.
	\end{split}
\end{equation}
The matrix is non-zero only in the non-compact directions, i.e.~along the generators $t_{\alpha} \in {\mathfrak e}_{6(6)} \setminus \mathfrak{usp}(8)$.
Moreover all goldstone fields appear with a zero eigenvalue.


\section{Extrema of the scalar potential} 
\label{sec:extrema_of_the_scalar_potential}

The procedure used to find and analyze the scalar potential has been developed in the case of maximal supergravities in \cite{DallAgata:2011aa}, developing on an old idea presented in a very different context \cite{Li:1986tk}.
The main point is that the scalar potential is a function of the scalar fields via the coset representatives ${\cal V}_M^{ij}$ and the embedding tensor $\Theta_M{}^\alpha$
\begin{equation}
		V(\phi) = V\left({\cal V}(\phi),\Theta\right).
\end{equation}
As explained above, vacua of the theory follow as solutions of the minimization condition (\ref{eom}).
This is generally a rather complicated expression of the scalar fields (at best ratios of polynomials and exponentials of the scalar fields).
This is the reason why the task of finding solutions to such complicated system of equations has always been very challenging and researchers usually focussed on restricted sets of scalar fields in order to simplify the task, which anyway is often performed only numerically.

The alternative proposed in \cite{DallAgata:2011aa} maps the problem to a coupled set of second and first order algebraic conditions on the gauging parameters.
This is possible because the scalar manifold is homogeneous and therefore each point on the manifold can be mapped to any other by an E$_{6(6)}$ transformation and at the same time the scalar potential is \emph{invariant} under the simultaneous  action of these trasformations on both the coset representatives and on the embedding tensor.
This implies that we can always map any critical point of the scalar potential to the ``origin'' at $\phi = 0$.
At such point, the scalar potential is a simple quadratic function of the embedding tensor
\begin{equation}	
	V = \frac{2}{15}\, \Theta_M{}^\alpha \Theta_M{}^\beta\left(\delta_{\alpha \beta} + 5 \eta_{\alpha \beta}\right)
\end{equation}
and the minimization conditions become quadratic conditions on the embedding tensor, which should be solved together with the quadratic constraints (\ref{quadconst}).
The result is that rather than fixing the gauging and then performing a scan of all possible critical points of the scalar potential and then scan among all possible gaugings, one can simply solve a set of quadratic conditions on the embedding tensor and then read the resulting values of $\Theta$ that fix at the same time the gauge group, the value of the cosmological constant and the masses at the critical point.
Clearly any choice of point on the scalar manifold is equivalent, but choosing $\phi = 0$ has the advantage that it is a fixed point under the action of the maximal compact subgroup of the isometries, namely USp(8), and therefore we can consider modifications of the embedding tensor related only to the non-compact transformations, so that that there is a one-to-one correspondence between the parameters in $\Theta$ related to the scalar fields and the independent directions on the scalar manifold.
We advise the reader to consult \cite{DallAgata:2011aa} for more details.

As we mentioned in the introduction, all our results are fully analytic.
The reason we are able to produce such results is related to the procedure we used to solve the quadratic conditions coming from the minimization of the scalar potential and from the quadratic constraints.
While in fact we reduced our problem to a set of quadratic equations, we still have generically a very large number of parameters and quadratic equations.
This implies that not always one can see a straightforward analytic solution, because the equations are coupled and they could become very high in order in terms of a single variable.

We mainly used two techniques.
The first one is based on a simplification of the set of quadratic equations by employing a choice of a more convenient Gr\"obner basis for the polynomial generating the same solutions.
This has been done with the aid of the computer algebra system for polynomial computations {\tt SINGULAR} \cite{DGPS}.
Unfortunately when the number of variables is very large, this can be extremely costly in time and therefore one has to resort to a different way of reducing the set of equations.
We found a very effective procedure by borrowing an algorithm developed in the context of cryptography where the solution of quadratic equations on finite fields is a common problem.
In particular we used the so-called XL algorithm \cite{linearization}, or extended linearization.
The idea is rather simple.
Rather than solving directly the given set of quadratric equations, one produces sets of linear equations in the monomials appearing in the equations and in all equations obtained by multiplying the original set of equations by the variables and by their products up to a fixed order.
This produces sets of linear equations that can be solved rapidly and, once interpreted in terms of the original variables, they may reduce to equations in a single variable or in simpler sets of polynomial equations (like equality between different monomials).
This allows to fix and eliminate some of the variables from the problem and then face a simpler set of equations, which could be solved directly or further simplified by another iteration of the same procedure, or by a more convenient choice of Gr\"obner basis.


\section{Vacua with residual U(2) symmetry} 
\label{sec:vacua_with_residual_u_2_symmetry}

In this work we decided to scan gauged maximal supergravity in 5 dimensions for vacua with a residual U(2) symmetry.
Asking for a residual U(2) invariance of the vacuum (with respect to a gauged or global symmetry) imposes restrictions on the allowed coefficients of the embedding tensor and consequently of the fermion shift tensors, which should be singlets with respect to this residual symmetry.
To perform a full analysis, we therefore looked at all the inequivalent embeddings of SU(2) in USp(8) and then singled out all possible inequivalent charge assignments for the remaining U(1)s, if any.
We then performed the branching of the $\mathbf{36}$ and $\mathbf{315}$ representations of USp(8) specifying the fermion shifts with respect to the chosen embedding and classified all inequivalent cases.
When the commutant of the residual symmetry group in USp(8) was non-trivial we used the commuting symmetries to further reduce the number of inequivalent variables by removing those that could be generated by the action of the commutant.
Once the non-vanishing components of the fermion shifts had been identified we then proceeded to solve the set of quadratic algebraic conditions coming from the scalar equations of motion (\ref{eom}) and the quadratic constraints (\ref{quadconst}) and then collected all solutions, which may still be related by duality transformations.
Finally, we analyzed their properties and computed their mass spectrum as we will discuss momentarily.
In the summary tables we collected all inequivalent vacua and reported the most general mass spectra for each of them.
Unfortunately, two of the branchings still present a very large number of singlets ($\geq 48$) and even combining all the techniques mentioned above we have not been able to fully scan and solve their equations for all the allowed parameters, though for all solutions we recovered the same vacua we found in other branchings.

As a first step we list the branchings we analyzed by the inequivalent decompositions of the \textbf{8}-dimensional representation of USp(8) under SU(2) and then give one of the branching routes leading to this decomposition.
For each case we also give a table with the subcases based on possible different choices of the U(1) factor, when present.
We also list the number of singlets in the fermion shifts, which are going to be the variables to be fixed by the quadratic conditions in order to find vacua.

\subsection{Branchings} 
\label{sub:branchings}

We find 13 different branchings of the fundamental representation of USp(8) under SU(2), which we therefore analyze separately.
The labels on the various factors are self explanatory: we use letters from the beginning of the alphabet to keep track of the various factors in the decompositions and we use $S$ and $diag$ to specify the symmetric and diagonal embedding of the group.

\paragraph{Case 1: $\mathbf{8} \to \mathbf{8}$.\newline}

The branching path is
\begin{equation}
	{\rm USp}(8) \to {\rm SU}(2)_S.
\end{equation}
This case leaves no singlets to discuss, so no vacua are possible for this choice.

\paragraph{Case 2: $\mathbf{8} \to \mathbf{6}+ \mathbf{2}$.\newline} 

The branching path is
\begin{equation}
	{\rm USp}(8) \to {\rm SU}(2)_A \times {\rm USp}(6) \to  {\rm SU}(2)_A \times {\rm SU}(2)_S \to {\rm SU}(2)_{diag} 
\end{equation}
There is only one singlet in $A_{i,jkl}$.

\paragraph{Case 3: $\mathbf{8} \to \mathbf{6}+ \mathbf{1} + \mathbf{1}$.\newline} 

The branching path is
\begin{equation}
	{\rm USp}(8) \to {\rm SU}(2)_A \times {\rm USp}(6) \to  1_A \times {\rm SU}(2)_S 
\end{equation}
There are 3 singlets in $A_{ij}$ and no singlets in $A_{i,jkl}$. 

\paragraph{Case 4: $\mathbf{8} \to \mathbf{4}+ \mathbf{4}$.\newline} 
The branching path is
\begin{equation}
	{\rm USp}(8) \to {\rm USp}(4)^2 \to {\rm USp}(4)_{diag} \to SU(2)_S
\end{equation}
We have one singlet in $A_{ij}$ and 3 singlets in $A_{i,jkl}$.


\paragraph{Case 5: $\mathbf{8} \to \mathbf{4}+ \mathbf{2} + \mathbf{2}$.\newline} 

The branching path is
\begin{equation}
	{\rm USp}(8) \to {\rm USp}(4)_A \times {\rm USp}(4)_B \to [{\rm SU}(2) \times {\rm U}(1)]_{A} \times {\rm SU}(2)_S \to {\rm SU}(2)_{diag}
\end{equation}
We find just one singlet in $A_{ij}$ and 8 in ${A_{i,jkl}}$.

\paragraph{Case 6: $\mathbf{8} \to \mathbf{4}+ \mathbf{2} + \mathbf{1} + \mathbf{1}$.\newline} 

The branching path is
\begin{equation}
	{\rm USp}(8) \to {\rm USp}(4) \times {\rm USp}(4) \to [{\rm SU}(2)_A \times {\rm SU}(2)_B] \times {\rm SU}(2)_S \to {\rm SU}(2)_{B+S}
\end{equation}
There are 3 singlets in $A_{ij}$ and 5 singlets in $A_{i,jkl}$.

\paragraph{Case 7: $\mathbf{8} \to \mathbf{4}+ 4\cdot \mathbf{1}$.\newline}
The branching path is
\begin{equation}
	{\rm USp}(8) \to {\rm USp}(4)_A \times {\rm USp}(4)_B \to [{\rm SU}(2)_S]_A \times [{\rm SU}(2) \times {\rm U}(1)]_{B} \to {\rm SU}(2)_{S} \times {\rm U}(1)_A \times {\rm U}(1)_B
\end{equation}

In this case we have two inequivalent choices of U(1) $\subset$ U(1)$_A \times$ U(1)$_B$, which we list in the table 1.
\begin{table}[H]
\begin{center}
	\rowcolors{1}{gray!15}{white}
\begin{tabular}{|c|cccc|}
\hline
\rowcolor{white}\#  & \textbf{8}  & charges  & \textbf{36}  &  \textbf{315}  \\[2mm]
 & decomposition &choice &  singlets & singlets\\[2mm]
\hline\hline
7 &$\mathbf{8} \to \mathbf{4}_{00} + \mathbf{1}_{\pm1\pm1}$& $(q_A,q_B)$&& \\[2mm]
7a & $\textbf{8} \to \mathbf{4}_0 + 2\cdot \mathbf{1}_{\pm 1}$& $q_A$& 4& 7\\[2mm]
7b & $\textbf{8} \to \mathbf{4}_0 + \mathbf{1}_{\pm 1}+ 2 \cdot \mathbf{1}_{0}$& $\frac{q_A+q_B}{2}$& 4& 5\\[2mm]
\hline
\end{tabular}\label{tablecase7}
\caption{Branchings for the case 7.}
\end{center}
\end{table}


\paragraph{Case 8: $\mathbf{8} \to 2 \cdot \mathbf{3}+  \mathbf{2}$.\newline}
The branching path is
\begin{equation}
	{\rm USp}(8) \to {\rm SU}(2)_A \times {\rm USp}(6) \to  {\rm SU}(2)_A  \times [{\rm SU}(3) \times {\rm U}(1)]_B \to {\rm SU}(2)_A  \times {\rm SO}(3)_B \to {\rm SU}(2)_{diag}
\end{equation}
The decomposition contains 3 singlets for $A_{ij}$ and 6 singlets for $A_{i,jkl}$.

\paragraph{Case 9: $\mathbf{8} \to 2 \cdot \mathbf{3}+  2\cdot \mathbf{1}$.\newline}
The branching path is
\begin{equation}
	{\rm USp}(8) \to {\rm SU}(4) \times {\rm U}(1)_B\to {\rm SU}(3) \times {\rm U}(1)_A \times {\rm U}(1)_B\to {\rm SU}(2)_{s} \times {\rm U}(1)_A \times {\rm U}(1)_B.
\end{equation}
This case has already 21 SU(2) singlets overall, therefore we distinguish various subcases according to the choices of a U(1) factor, which we report in the table 2.
\begin{table}[H]
\begin{center}
	\rowcolors{1}{gray!15}{white}
\begin{tabular}{|c|cccc|}
\hline
\rowcolor{white}\#  & \textbf{8}  & charges  & \textbf{36}  &  \textbf{315}   \\[2mm]
 & decomposition &choice &  singlets & singlets\\[2mm]
\hline\hline
9 &$\mathbf{8} \to \mathbf{3}_{11} + \mathbf{3}_{-1-1} + \mathbf{1}_{-31}+ \mathbf{1}_{3-1}$&$(q_A,q_B)$&2&1 \\[2mm] 
9a & $\mathbf{8} \to \mathbf{3}_{\pm1} + \mathbf{1}_{\pm3}$&$q_A$&2&3 \\[2mm] 
9b & $\mathbf{8} \to \mathbf{3}_{\pm1} + \mathbf{1}_{\pm1}$&$q_B$, $\frac{q_A+q_B}{2}$&2&5 \\[2mm] 
9c & $\mathbf{8} \to 2\cdot \mathbf{3}_{0} + \mathbf{1}_{\pm 1}$&$\frac{q_A-q_B}{4}$&4&3 \\[2mm] 
9d & $\mathbf{8} \to \mathbf{3}_{\pm1} + 2\cdot\mathbf{1}_{0}$&$\frac{q_A+3q_B}{4}$&4&1 \\[2mm] 
\hline
\end{tabular}\label{tablecase9}
\caption{Branchings for the case 9.}
\end{center}
\end{table}


\paragraph{Case 10: $\mathbf{8} \to 4\cdot \mathbf{2}$.\newline}

The branching path is
\begin{equation}
	{\rm USp}(8) \to {\rm SU}(2)_A \times {\rm SU}(2)_B \times {\rm SU}(2)_C \to {\rm SU}(2)_{C} \times {\rm U}(1)_A \times {\rm U}(1)_B
\end{equation}
This case has 51 singlets of SU(2) and therefore we classify various subcases according to a remaining U(2) symmetry.
We collect all different branchings in table 3.
\begin{table}[H]
\begin{center}
	\rowcolors{1}{gray!15}{white}
\begin{tabular}{|c|cccc|}
\hline
\rowcolor{white}\#  & \textbf{8}  & charges  & \textbf{36}  &  \textbf{315}   \\[2mm]
 & decomposition &choice &  singlets & singlets\\[2mm]
\hline\hline
10 & $\mathbf{8} \to \mathbf{2}_{\pm 1 \pm 1}$& $(q_A,q_B)$&2&3 \\[2mm]
10a& $\textbf{8} \to 2\cdot[\mathbf{2}_{\pm 1}]$ & $q_A$ & 4 & 15 \\[2mm]
10b& $\textbf{8} \to 2\cdot[\mathbf{2}_{0}] + 2_{\pm 1}$ & $\frac{q_A+q_B}{2}$ & 2 & 11 \\[2mm]
10c& $\textbf{8} \to [\mathbf{2}_{\pm 3}] + [\mathbf{2}_{\pm 1}]$ & $2q_A+q_B$ & 2 & 7 \\[2mm]  \hline
\end{tabular}
\end{center}\label{tablecase10}
\caption{Branchings for the case 10.}
\end{table}

\paragraph{Case 11: $\mathbf{8} \to 3\cdot \mathbf{2} + 2 \cdot \mathbf{1}$.\newline}
The branching path is
\begin{equation}
	\begin{split}
	{\rm USp}(8) &\to {\rm SU}(2) \times {\rm USp}(6) \to  {\rm SU}(2) \times [{\rm SU}(3) \times {\rm U}(1)] \\[2mm]
	&\to {\rm SU}(2) \times [{\rm SU}(2) \times {\rm U}(1) \times {\rm U}(1)] \to {\rm SU}(2)_{diag} \times {\rm U}(1)_A \times {\rm U}(1)_B
	\end{split}
\end{equation}
This case has 39 singlets of SU(2) and therefore we classify various subcases according to a remaining U(2) symmetry.
Results are collected in table 4.
\begin{table}[H]
\begin{center}
	\rowcolors{1}{gray!15}{white}
\begin{tabular}{|c|cccc|}
\hline
\rowcolor{white}\#  & \textbf{8}  & charges  & \textbf{36}  &  \textbf{315}   \\[2mm]
 & decomposition &choice &  singlets & singlets\\[2mm]
\hline\hline
11 & $\mathbf{8} \to \mathbf{2}_{11} + \mathbf{2}_{00} + \mathbf{2}_{-1-1} + \mathbf{1}_{2-1} + \mathbf{1}_{-21}$& $(q_A,q_B)$&2&5 \\[2mm]
11a& $\textbf{8} \to \mathbf{2}_{\pm1} + \mathbf{2}_0 + 1_{\pm2}$ & $q_A$ & 2 & 5 \\[2mm]
11b& $\textbf{8} \to \mathbf{2}_{\pm1} + \mathbf{2}_0 + 1_{\pm1}$ & $q_B$ & 2 & 7 \\[2mm]
11c& $\textbf{8} \to \mathbf{2}_{\pm2} + \mathbf{2}_0 + 1_{\pm1}$ & $q_A+q_B$ & 2 & 7 \\[2mm]
11d& $\textbf{8} \to 3\cdot \mathbf{2}_0 + 1_{\pm1}$ & $\frac{q_A-q_B}{3}$ & 4 & 23 \\[2mm]
11e& $\textbf{8} \to \mathbf{2}_{\pm1} + \mathbf{2}_0  + 2 \cdot 1_{0}$ & $\frac{q_A+2q_B}{3}$ & 4 & 7 \\[2mm]\hline
\end{tabular}
\end{center}\label{tablecase11}
\caption{Branchings for the case 11.}
\end{table}

\paragraph{Case 12: $\mathbf{8} \to 2\cdot \mathbf{2} + 4 \cdot \mathbf{1}$.\newline}
The branching path is
\begin{equation}
	\begin{split}
	{\rm USp}(8) &\to {\rm USp}(4)_A \times {\rm USp}(4)_B \to  [{\rm SU}(2) \times {\rm U}(1)]_A \times [{\rm SU}(2) \times {\rm SU}(2)]_B \\[2mm]
	&\to {\rm SU}(2)_A \times {\rm U}(1)_A  \times {\rm U}(1)_B \times {\rm U}(1)_C
	\end{split}
\end{equation}
This case has 64 singlets of SU(2) and therefore we classify various subcases according to a remaining U(2) symmetry, which we list in table 5.
\begin{table}[H]
\begin{center}
	\rowcolors{1}{gray!15}{white}
\begin{tabular}{|c|cccc|}
\hline
\rowcolor{white}\#  & \textbf{8}  & charges  & \textbf{36}  &  \textbf{315}   \\[2mm]
 & decomposition &choice &  singlets & singlets\\[2mm]
\hline\hline
12 &$\mathbf{8} \to \mathbf{2}_{\pm1 0 0} + \mathbf{1}_{0\pm10} + \mathbf{1}_{00\pm1}$& $(q_A,q_B,q_C)$&3&5 \\[2mm]
12a&$\mathbf{8} \to \mathbf{2}_{\pm1} + 4\cdot \mathbf{1}_{0}$&$q_A$&11&21\\[2mm]
12b&$\mathbf{8} \to 2 \cdot \mathbf{2}_{0} +  \mathbf{1}_{\pm1} + 2 \cdot \mathbf{1}_0$&$q_B$&5&19\\[2mm]
12c&$\mathbf{8} \to \mathbf{2}_{\pm1} + 2 \cdot \mathbf{1}_{\pm1}$&$q_A+q_B+q_C$&5&19\\[2mm]
12d&$\mathbf{8} \to \mathbf{2}_{\pm1} + \mathbf{1}_{\pm1}+ 2 \cdot \mathbf{1}_0$&$q_A+q_B$&5&9\\[2mm]
12e&$\mathbf{8} \to 2\cdot[\mathbf{2}_{0} + \mathbf{1}_{\pm1}]$&$q_B+q_C$&5&27\\[2mm]
12f&$\mathbf{8} \to \mathbf{2}_{\pm1} + \mathbf{1}_{\pm2} + 2 \cdot \mathbf{1}_0$&$q_A+2q_B$&5&15\\[2mm]\hline
\end{tabular}
\end{center}\label{tablecase12}
\caption{Branchings for the case 12.}
\end{table}

\paragraph{Case 13: $\mathbf{8} \to \mathbf{2} + 6 \cdot \mathbf{1}$.\newline}
The branching path is
\begin{equation}
	\begin{split}
	{\rm USp}(8) &\to {\rm SU}(2) \times {\rm USp}(6) \to  {\rm SU}(2) \times [{\rm SU}(2)_A \times {\rm USp}(4)] \\[2mm]
	&\to {\rm SU}(2) \times [{\rm SU}(2)_A \times {\rm SU}(2)_B \times {\rm SU}(2)_C] \to   {\rm SU}(2) \times {\rm U}(1)_A \times {\rm U}(1)_B \times {\rm U}(1)_C
	\end{split}
\end{equation}
This case has 124 singlets of SU(2) and therefore we classify various subcases according to a remaining U(2) symmetry.
Note that cases 13e and 13f have only a subset of the singlets present in the other cases, so it is enough to solve cases 13a--13d.
Results are presented in table 6.

The branchings 13a and 13d present more than 48 singlets and this hampered the simplification of the problem with any of the techniques used in this work in a reasonable amount of time.
Anyway, all solutions we have been able to find for these branchings were already present in one of the other branchings.
\begin{table}[H]
\begin{center}
	\rowcolors{1}{gray!15}{white}
\begin{tabular}{|c|cccc|}
\hline
\rowcolor{white}\#  & \textbf{8}  & charges  & \textbf{36}  &  \textbf{315}   \\[2mm]
 & decomposition &choice &  singlets & singlets\\[2mm]
\hline\hline
13 &$\mathbf{8} \to \mathbf{2}_{000} + \mathbf{1}_{\pm100}+ \mathbf{1}_{0\pm10}+ \mathbf{1}_{00\pm1}$&$(q_A,q_B,q_C)$&3&9\\[2mm]
13a*& $\mathbf{8} \to \mathbf{2}_{0} + \mathbf{1}_{\pm1} + 4\cdot \mathbf{1}_{0}$& $q_A$ & 11 & 37 \\[2mm]
13b&$\mathbf{8} \to \mathbf{2}_{0}+ 2 \cdot[\mathbf{1}_{0} + \mathbf{1}_{\pm1}]$& $q_A+q_B$&7&27\\[2mm]
13c&$\mathbf{8} \to \mathbf{2}_{0}+ \mathbf{1}_{\pm2} + \mathbf{1}_{\pm1} + 2\cdot \mathbf{1}_{0}$&$2q_A+q_B$&5&17\\[2mm]
13d*& $\mathbf{8} \to \mathbf{2}_{0} + 3\cdot \mathbf{1}_{\pm1}$& $q_A+q_B+q_C$ & 9 & 47 \\[2mm]
13e& $\mathbf{8} \to \mathbf{2}_{0} + 2 \cdot \mathbf{1}_{\pm1}+ \mathbf{1}_{\pm2}$& $q_A+q_B+2q_C$ & 3 & 9 \\[2mm]
13f& $\mathbf{8} \to \mathbf{2}_{0} + \mathbf{1}_{\pm1}+2\cdot \mathbf{1}_{\pm2}$& $q_A+2q_B+2q_C$ & 3 & 19 \\[2mm]\hline
\end{tabular}
\end{center}\label{tablecase13}
\caption{Branchings for the case 13.}
\end{table}


\subsection{Vacua} 
\label{sub:vacua}

The search for vacua has been carried out by solving the sets of quadratic equations for the singlets in the tables above.
Once we found solutions, we checked for each candidate vacuum the rank of the embedding tensor, the signature of the resulting Cartan--Killing matrix and the full mass spectrum.
Overall we found 5 different Anti-de Sitter vacua, 5 Minkowski vacua and 2 de Sitter vacua.
The vacua with negative cosmological constant are all pertaining to the same gauging, namely the maximal SO(6) theory of \cite{Pernici:1985ju}, and were all already known \cite{Girardello:1998pd,Distler:1998gb,Khavaev:1998fb}.
Among the Minkowski vacua there are the Cremmer--Scherk--Schwarz gaugings \cite{Scherk:1979zr,Cremmer:1979uq} with various mass parameters and a supersymmetric vacuum for the SO$^*$(6) theory discovered in \cite{Gunaydin:1985tb}, but we also find three new vacua with a non-abelian gauge group (like those in \cite{Catino:2013ppa} for the analogous analysis of maximal supergravity in 4 dimensions).
Finally, we also find 2 de Sitter vacua, resulting from gauging of the semisimple groups SO(3,3) \cite{Gunaydin:1985cu} and SO$^*(6)$, the latter being new.
All the vacua are reported in the table~7, together with the number of supersymmetry they preserve, the original gauging, the residual gauge group and the reference where they were first discovered.
In the appendix we provide for each vacuum one instance of fermion shift values reproducing the critical point mentioned in the table.

\begin{table}[H]
\begin{center}
\rowcolors{1}{white}{gray!15}
\begin{tabular}{|c|ccccc|}
\hline
vacuum & susy & G$_{gauge}$ & G$_{res}$ & ref. & branching\\[2mm]
\hline\hline
A1 & 8 & SO(6) & SO(6) & \cite{Pernici:1985ju,Khavaev:1998fb}  & 4,9,10,12\\[2mm]
A2 & 0 & SO(6) & SO(5) & \cite{Girardello:1998pd,Distler:1998gb,Khavaev:1998fb} & 4, 10, $12_{cef}$\\[2mm]
A3 & 0 & SO(6) & SU(3) & \cite{Girardello:1998pd,Distler:1998gb,Khavaev:1998fb} & $9_a$, $12_{ef}$\\[2mm]
A4 & 2 & SO(6) & SU(2) $\times$ U(1) & \cite{Khavaev:1998fb} & $12_{cef}$\\[2mm]
A5 & 0 & SO(6) & SU(2) $\times$ U(1) $\times$ U(1) & \cite{Khavaev:1998fb} & $12_{be}$\\[2mm]
M1 & 0,2,4,6 & U(1)$ \ltimes {\mathbb R}^{16}$ & U(1) & \cite{Cremmer:1979uq} & \begin{tabular}{c}
5,6,7,10 \\
11,12,13 \\
\end{tabular}\\[2mm]
M2 & 2 & SO$^*$(6)=SU(3,1) & SU(3) $\times$ U(1) & \cite{Gunaydin:1985tb} & \begin{tabular}{c}
8,9,11\\
$12_{abcef}$, $13_b$ \\
\end{tabular} \\[2mm]
M3 & 4 & SO$^*$(4) $\ltimes {\mathbb R}^{8}$& U(2)  & here  & \begin{tabular}{c}
$12_{abcef}$ \\
11, $13_{bc}$\\
\end{tabular}\\[2mm]
M4 & 0 & [SO(3,1) $\times$ SO(2,1)] $\ltimes {\mathbb R}^{8}$ & U(2) & here &$10_b$ \\[2mm]
M5 & 4 & SO*(4) $\ltimes {\mathbb R}^{8}$ & SO(3) & here &$12_{be}$\\[2mm]
D1 & 0 & SO(3,3) & SO(3)$^2$ & \cite{Gunaydin:1985cu} &$9_b$,$10_{ab}$ \\[2mm]
D2 & 0 & SO$^*$(6)=SU(3,1)   & SU(2) & here &$9_b$ \\[2mm]
\hline
\end{tabular}
\end{center}
\label{tab:summary}
\caption{Summary of vacua found in this work.}
\end{table}

Given the nature of the gaugings generating such vacua, we can also see how some of these could be obtained from string theory reductions.
All AdS vacua appear in the SO(6) theory, which is a consistent truncation of type IIB supergravity compactified on $S^5$ \cite{Hohm:2014qga}.
A subset of the CSS gaugings and their vacua M1 are known to be the result of a twisted torus reduction \cite{Scherk:1979zr}, while the most general gauging and vacuum in this class may admit an uplift through a generalised Scherk--Schwarz Ansatz analogous to the ones described for four-dimensional CSS gaugings in \cite{Inverso:2019mnv}.

It is interesting to notice that for the first time in a maximal theory in $D \ge 4$ we find a gauging that produces at the same time vacua with different types of cosmological constants.
This is the SO$^*(6) = $SU(3,1) gauging, that contains at the same time a Minkowski and a de Sitter vacuum.
Our claim that they reside in the same model follows both from the analysis of the embedding tensors that generate them, and the direct identification of a truncated scalar potential for the SU(3,1) theory where both vacua are easily found.
From the embedding tensors we find which generators of ${\mathfrak e}_{6(6)}$ are involved in their corresponding model and analyzing the commutants we find in both cases that the representation \textbf{27} decomposes in the representations $\mathbf{15} + \mathbf{6} + \mathbf{6}$ of the gauge group.
This corresponds to the correct branching under SU(3,1) and since the adjoint is unique in the branching, we argue that the gaugings are the same.
Moreover, if we directly decompose the \textbf{315} representation of ${\mathfrak e}_{6(6)}$ from the branching above for the \textbf{27} we see that there is a unique singlet with respect to SU(3,1) and therefore there is a unique possible form of embedding tensor leading to this gauging up to duality transformations.

\begin{figure*}[ht]
	\begin{center}
 \includegraphics[scale=.4]{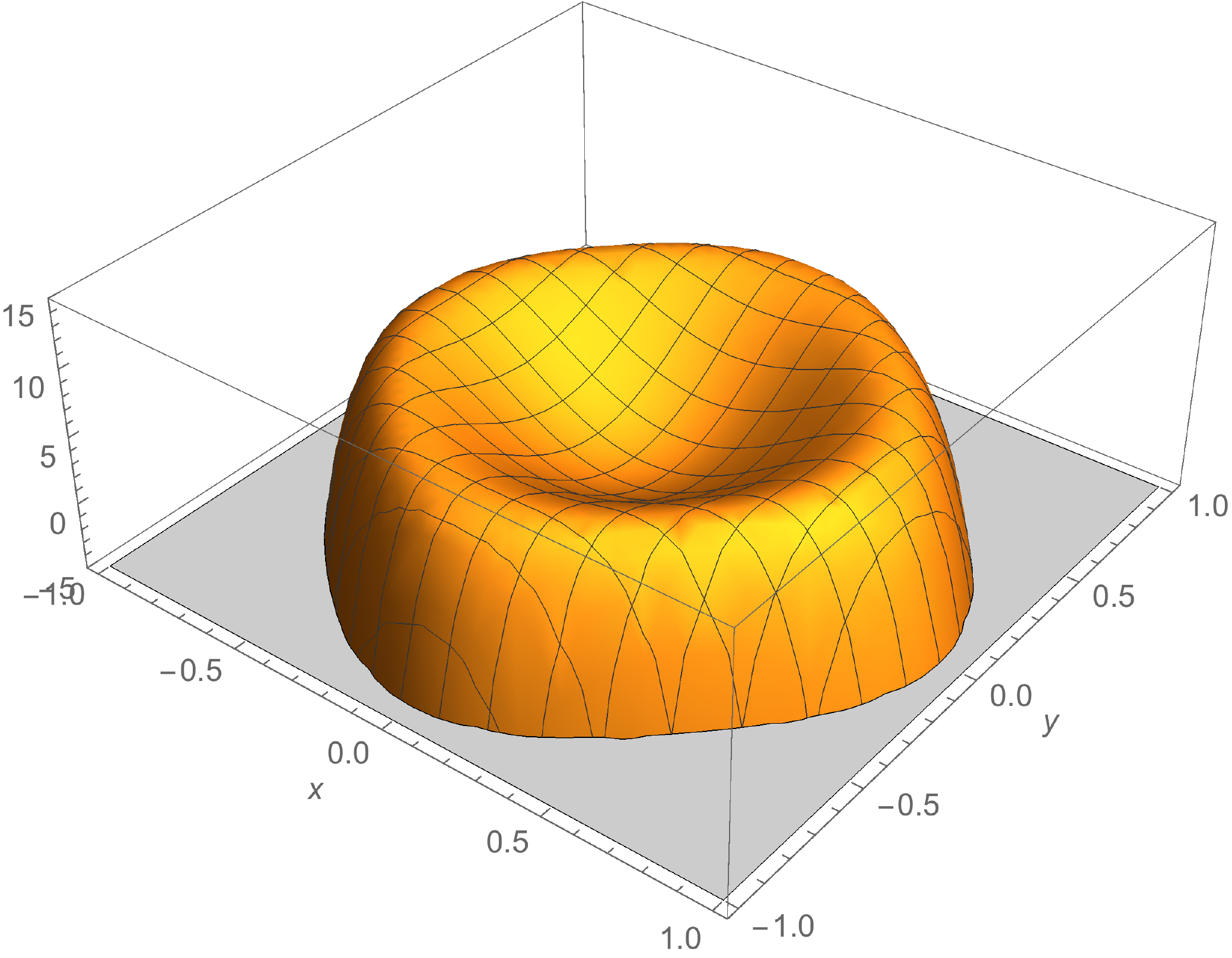} 
 	\end{center}
\caption{\label{fig:potential} Scalar potential for the two common scalars invariant under the residual symmetries of the vacua (M2) and (D2). We see a Minkowski vacuum at the center of the picture, surrounded by a family of de Sitter vacua, with a massless modulus.}
\end{figure*}
Actually, for this specific model we can provide a truncated scalar potential, where we make explicit the dependence on the two scalar fields that are singlets of both symmetry groups.
Furthermore, both vacua arise as different solutions of the $9_b$ case and the commutator of the residual U(2) group with the non-compact generators of $\mathfrak{e}_6$ leaves only two generators $g_1$ and $g_2$, for which we can provide a truncated scalar potential where both vacua can be found.
We construct the coset representative
\begin{equation}
	L(x, y) = \exp\left(g_1\, x + \,g_2\, y\right),
\end{equation}
which induces the scalar potential
\begin{equation}
		V = -\frac{27}{16}\left(12 - 16\cosh(2x) \cosh(2y) +4 \cosh^2(2x) \cosh^2(2y) \right),
\end{equation}
where $x$ and $y$ are canonically normalized scalar fields.
The scalar potential has two vacua, a Minkowski one at $x=y=0$ and a line of unstable de Sitter vacua at $\cosh(2x)\cosh(2y) = 2$.
At any point in the family of de Sitter vacua we see that the masses of the two fluctuations are indeed zero and $m^2/\Lambda = -24$.
These coincide with one of the moduli and one of the unstable directions of the full scalar spectrum about the de Sitter vacuum (see table 14).

A similar discussion could apply to the vacua (M3) and (M5).
They both have the same gauge group, though in this case they do not belong to the same model.
In fact, there are 4 U(2) invariant scalar fields in both models, but the scalar potentials show only a single vacua in each of the potentials constructed from (M3) and (M5) by introducing the appropriate coset representatives.
For instance, using canonically normalized fields, the potential of (M3) is 
\begin{equation}
	\begin{split}
		V = &\frac{x_1^{-\frac{4}{\sqrt{3}}}x_3^{-3 \sqrt{2}}}{8192\, x_2^2 x_4^2}  \left(x_3^{\sqrt{2}}-1\right)^2 \left[-8 \,m_1 m_2 \,x_1^{\sqrt{3}}\,x_2\, \left(x_2^2-1\right) \left(3 x_3^{2\sqrt{2}}+2 x_3^{\sqrt{2}}+3\right)^2 \left(x_4^4-1\right) \right.\\[2mm]
		 &+m_1^2 \,x_1^{2 \sqrt{3}}\left(\left(3 x_3^{2\sqrt{2}}+2 x_3^{\sqrt{2}}+3\right) \left(x_4^2+1\right) \left(1+x_2^2\right)+4 \left(x_3^{\sqrt{2}}-1\right)^2 x_4 x_2 \right)\\[2mm]
		 &\left(\left(3 x_3^{2\sqrt{2}}+2 x_3^{\sqrt{2}}+3\right) \left(x_4^2+1\right) \left(1+x_2^2\right)-4 \left(x_3^{2\sqrt{2}}+6x_3^{\sqrt2}+1\right) x_4 x_2 \right)\\[2mm]
		&+m_2^2 \left(\left(3 x_3^{2\sqrt{2}}+2 x_3^{\sqrt{2}}+3\right) \left(x_4^2+1\right) \left(1+x_2^2\right)-4 \left(x_3^{\sqrt{2}}-1\right)^2 x_4 x_2 \right) \\[2mm]
		&\left.\left(\left(3 x_3^{2\sqrt{2}}+2 x_3^{\sqrt{2}}+3\right) \left(x_4^2+1\right) \left(1+x_2^2\right)+4 \left(x_3^{2\sqrt{2}}+6x_3^{\sqrt2}+1\right) x_4 x_2 \right)\right].
	\end{split}
\end{equation}
This shows a single critical point at $x_i=1$, where the scalars $x_{1,2,4}$ are moduli, while the scalar $x_3$ is massive with mass $m_2$.
Actually $x_1$ is a modulus that simply rescales the mass parameters.
While the gauge group is the same, the two vacua indeed pertain to two different gaugings.
This is possible because the decomposition of the \textbf{351} of E$_{6(6)}$ under SO*(4) shows 6 singlets and therefore one could find inequivalent embeddings of the same gauge group.


\subsection{Mass spectra} 
\label{sub:mass_spectra}

In this final section we present the mass spectra of all the vacua listed in the previous table.
The masses for backgrounds with non-vanishing cosmological constant are normalized in terms of the (A)dS radius squared $L^2 = |6/V|$ , so that supersymmetric gravitinos have a normalized squared mass of $9/4$.

For the AdS vacua, which are not new, most of these spectra were already known from previous work, though the spectrum of the non-supersymmetric ones (A2), (A3) and (A5) was lacking some states that we provide.

\hspace{-1.2cm}
\begin{tabular}{cc}
\begin{minipage}{.5\linewidth}
\begin{table}[H]
\rowcolors{1}{white}{gray!15}
\begin{tabular}{|c|c|}
	\hline
	$L^2 m^2_{3/2}$ & $\left[\frac94\right]_{\times 8}$  \\[2mm]
	$L^2 m^2_{vec}$ & $[0]_{\times 15}$  \\[2mm]
	$L^2 m^2_{tens}$ &   $[1]_{\times 12}$  \\[2mm]
	$L^2 m^2_{1/2}$ & $\left[\frac14\right]_{\times 40}$,  $\left[\frac94\right]_{\times 8}$ \\[2mm]
	$L^2 m^2_{scal}$ & $[-4]_{\times 20}$, $[-3]_{\times 20}$, $[0]_{\times 2}$ \\[2mm]
	\hline
\end{tabular}
\centering
\caption{Masses for the AdS vacuum A1}
\end{table}
\end{minipage}
&
\begin{minipage}{.5\linewidth}
\begin{table}[H]
\rowcolors{1}{white}{gray!15}
\begin{tabular}{|c|c|}
	\hline
	$L^2 m^2_{3/2}$ & $\left[\frac{8}{3}\right]_{\times 8}$  \\[2mm]
	$L^2 m^2_{vec}$ & $[0]_{\times 10}$, $\left[\frac83\right]_{\times 5}$  \\[2mm]
	$L^2 m^2_{tens}$ &   $\left[\frac23\right]_{\times 10}$, $[6]_2$  \\[2mm]
	$L^2 m^2_{1/2}$ & $[0]_{\times 32}$,  $\left[\frac{8}{3}\right]_{\times 8}$,  $\left[\frac{675}{128}\right]_{\times 8}$ \\[2mm]
	$L^2 m^2_{scal}$ & $\left[-\frac{16}{3}\right]_{\times 14}$, $[-2]_{\times 20}$,  $[0]_{\times 7}$, $[8]_{\times 1}$ \\[2mm]
	\hline
\end{tabular}
\centering
\caption{Masses for the AdS vacuum A2}
\end{table}
\end{minipage}
\end{tabular}

\begin{table}[H]
	\begin{center}
\rowcolors{1}{white}{gray!15}
\begin{tabular}{|c|c|}
	\hline
	$L^2 m^2_{3/2}$ & $\left[\frac{49}{18}\right]_{\times 6}$, $\left[\frac{9}{2}\right]_{\times 2}$ \\[2mm]
	$L^2 m^2_{vec}$ & $[0]_{\times 8}$, $\left[\frac{32}{9}\right]_{\times 6}$, \ $[8]_{\times 1}$\\[2mm]
	$L^2 m^2_{tens}$ & $\left[\frac89\right]_{\times 6}$, $\left[\frac{32}{9}\right]_{\times 6}$\\[2mm]
	$L^2 m^2_{1/2}$ & $[0]_{\times 8}, \ \left[\frac12\right]_{\times 16}$,   $\left[\frac{25}{18}\right]_{\times 18}$,   $\left[\frac{121}{18}\right]_{\times 6}$ \\[2mm]
	$L^2 m^2_{scal}$ & $\left[-\frac{40}{9}\right]_{\times 12}$, $\left[-\frac{16}{9}\right]_{\times 12}$, $[0]_{\times 17}$, $[8]_{\times 1}$ \\[2mm]
	\hline
\end{tabular}
\end{center}
\caption{Masses for the AdS vacuum A3}
\end{table}

The spectrum of the vacuum (A4) is particularly interesting in the context of the AdS/CFT correspondence, as it fixes the anomalous dimensions of the operators of the corresponding N=1 deformation of super-Yang--Mills in 4 dimensions \cite{Freedman:1999gp}.

\begin{table}[H]
	\begin{center}
\rowcolors{1}{white}{gray!15}
\begin{tabular}{|c|c|}
	\hline
	$L^2 m^2_{3/2}$ & $\left[\frac{49}{16}\right]_4$, \ $[4]_2$,\ $\left[\frac94\right]_{\times 2}$  \\[2mm]
	$L^2 m^2_{vec}$ & $[0]_4$, \ $\left[\frac{9}{16}\right]_{\times 4}$,\ $\left[\frac{5}{4}\right]_{\times 2}$, \ $\left[\frac{65}{16}\right]_{\times 4}$, \ $[6]_{\times 1}$ \\[2mm]
	$L^2 m^2_{tens}$ & $\left[\frac{9}{4}\right]_{\times 2}$, $\left[\frac{9}{16}\right]_{\times 4}$, \ $\left[\frac{25}{16}\right]_{\times 4}$, \ $\left[\frac{25}{4}\right]_{\times 2}$, \\[2mm]
	$L^2 m^2_{1/2}$ & $\begin{array}{c}
	\left[\frac{1}{16}\right]_{\times 4},\ \left[\frac{1}{4}\right]_{\times 6}, \ \left[\frac{9}{16}\right]_{\times 4},\ [1]_{\times 2}, \ \left[\frac{25}{16}\right]_{\times 4},\ \left[\frac{9}{4}\right]_{\times 2},  \\[2mm]
	\left[\frac{49}{16}\right]_{\times 8}, [4]_{\times 2}, \ [0]_{\times 12}, \  \left[\frac{29}{4}\pm \sqrt7\right]_{\times 2} \\[2mm]
	\end{array}$\\[2mm]
	$L^2 m^2_{scal}$ & $[0]_{\times 13}$, \ $[-4]_3$, \ $\left[-\frac{15}{4}\right]_{\times 12}$,\ $\left[-\frac{55}{16}\right]_{\times 4}$, \ $[-3]_{\times 2}$,\ $\left[-\frac{39}{16}\right]_{\times 4}$,\  $[3]_{\times 2}$, \ $[4\pm2 \sqrt7]_{\times 1}$ \\[2mm]
	\hline
\end{tabular}
\end{center}
\caption{Masses for the AdS vacuum A4}
\end{table}

\begin{table}[H]
	\begin{center}
\rowcolors{1}{white}{gray!15}
\begin{tabular}{|c|c|}
	\hline
	$L^2 m^2_{3/2}$ & $\left[\frac{81}{25}\right]_{\times 4}$, $\left[\frac{18}{5}\right]_{\times 4}$ \\[2mm]
	$L^2 m^2_{vec}$ & $\left[0\right]_{\times 4},\ \left[\frac{24}{5}\right]_{\times 1},\ \left[\frac{96}{25}\right]_{\times 8},\ \left[\frac{24}{25}\right]_{\times 2}$ \\[2mm]
	$L^2 m^2_{tens}$ & $\left[\frac{44}{5}\right]_{\times 2},\ \left[4\right]_{\times 2},\ \left[\frac{16}{25}\right]_{\times 8}$\\[2mm]
	$L^2 m^2_{1/2}$ &  $[0]_{\times 8},\ \left[\frac{22}{5}\pm 4 \sqrt{\frac25}\right]_{\times 4},\ \left[\frac{34}{25}\right]_{\times 8},\ \left[\frac{2}{5}\right]_{\times 4},\ \left[\frac{1}{25}\right]_{\times 12}$, $[\frac{161}{25}\pm \frac45\,\sqrt{34}]_{\times 4}$\\[2mm]
	$L^2 m^2_{scal}$ & $\left[\frac{52}{5}\right]_{\times 2},\ \left[\frac{84}{25}\right]_{\times 2},\ \left[\frac{48}{5}\right]_{\times 1},\ \left[-\frac{136}{25}\right]_{\times 6},\ \left[-4\right]_{\times 4},\ \left[-\frac{64}{25}\right]_{\times 8},\ \left[-\frac{12}{5}\right]_{\times 6},\ \left[0\right]_{\times 13}$ \\[2mm]
	\hline
\end{tabular}
\end{center}
\caption{Masses for the AdS vacuum A5. }
\end{table}

The full spectra of the de Sitter vacua (D1) and (D2) are new and show that such vacua are unstable with very large instabilities, of the order of the cosmological constant, or larger.

\hspace{-1.2cm}
\begin{tabular}{cc}
\begin{minipage}{.5\linewidth}
\begin{table}[H]
	\begin{center}
\rowcolors{1}{white}{gray!15}
\begin{tabular}{|c|c|}
	\hline
	$L^2 m^2_{3/2}$ & $[0]_{\times 8}$\\[2mm]
	$L^2 m^2_{vec}$ & $[0]_{\times 6}$, $[8]_{\times 9}$\\[2mm]
	$L^2 m^2_{tens}$ & $[2]_{\times 12}$\\[2mm]
	$L^2 m^2_{1/2}$ &  $[0]_{\times 16}$, $[8]_{\times 32}$\\[2mm]
	$L^2 m^2_{scal}$ & \begin{tabular}{c}
	$[-8]_{\times 1}$, $[-6]_{\times 2}$, \\[2mm]
	$[0]_{\times 11}$,    $[10]_{\times 18}$, $[16]_{\times 10}$ \\[2mm]
	\end{tabular}\\[2mm]
	\hline
\end{tabular}
\end{center}
\caption{Masses for the dS vacuum D1.}
\end{table}
\end{minipage}
\begin{minipage}{.5\linewidth}
\begin{table}[H]
	\begin{center}
\rowcolors{1}{white}{gray!15}
\begin{tabular}{|c|c|}
	\hline
	$L^2 m^2_{3/2}$ & $\left[\frac{9}{2}\right]_{\times 2}$, \ $\left[\frac{81}{2}\right]_{\times 6}$\\[2mm]
	$L^2 m^2_{vec}$ & $[0]_{\times 3}$, \    $[24]_{\times 1}$, \ $[96]_{\times 11}$\\[2mm]
	$L^2 m^2_{tens}$ & $[32]_{\times 6}$,\ $[56]_{\times 6}$\\[2mm]
	$L^2 m^2_{1/2}$ & \begin{tabular}{c}
	$[0]_{\times 8}$,\ $\left[\frac{25}{2}\right]_{\times 6}$,\ $\left[\frac{121}{2}\right]_{\times 10}$,\\[2mm]
	$\left[\frac{169}{2}\right]_{\times 6}$,\ $\left[\frac{225}{2}\right]_{\times 8}$, \  $\left[\frac{289}{2}\right]_{\times 10}$\ \\[2mm]
	\end{tabular} \\[2mm]
	$L^2 m^2_{scal}$ & \begin{tabular}{c}
	$[-24]_{\times 1}$, $[0]_{\times 14}$,\\[2mm]
	$[4(29\pm\sqrt{433})]_{\times 5}$,\\[2mm]
	$[40]_{\times 3}$, \  $[112]_{\times 12}$,  \   $[120]_{\times 2}$\\[2mm]
	\end{tabular}\\[2mm]
	\hline
\end{tabular}
\end{center}
\caption{Masses for the dS vacuum D2.}
\end{table}
\end{minipage}
\end{tabular}

\bigskip

\bigskip

For what concerns the Minkowski vacua, since there is no intrinsic scale associated to the vacuum, we parametrized all masses in terms of the ones of the gravitini.
We easily reproduce the expected spectrum for the CSS vacua, while the results for all the other vacua are new.

By also looking at the fermion shifts collected in the appendix, is interesting to notice that all the vacua we found show spectra that do not depend on additional parameters except for a few masses (or the cosmological constant, if different from zero).
This means that for all the gaugings considered the vacua appear in a unique theory with that gauge group and there are no continuous families of models with the same gauge group containing such vacua.
This differs from what was discovered in the 4-dimensional case \cite{DallAgata:2012mfj,DallAgata:2014tph}, where it was found that one can have infinite families of gaugings with the same gauge group and vacua whose existence and whose value of the cosmological constant may depend on the parameter specifying the family of gaugings.

\begin{table}[H]
	\begin{center}
\rowcolors{1}{white}{gray!15}
\begin{tabular}{|c|c|}
	\hline
	$m^2_{3/2}$ & $\left[m_1^2\right]_{\times 2},\ \left[m_2^2\right]_{\times 2},\ \left[m_3^2\right]_{\times 2},\ \left[m_4^2\right]_{\times 2}$\\[2mm]
	$m^2_{vec}$ & $[0]_{\times 1},\ [(m_1\pm m_3)^2]_{\times 2},\ [(m_1\pm m_4)^2]_{\times 2},\ [(m_2\pm m_3)^2]_{\times 2},\ [(m_2\pm m_4)^2]_{\times 2}$\\[2mm]
	$m^2_{tens}$ & $[0]_{\times 2},\ [(m_1\pm m_2)^2]_{\times 2},\ [(m_3\pm m_4)^2]_{\times 2}$\\[2mm]
	$m^2_{1/2}$ &  $\begin{array}{c}
	[0]_{\times 8},\ [m_i^2]_{\times 2},\ [(m_3 \pm m_1 \pm m_2)^2]_{\times 2},\ [(m_4 \pm m_1 \pm m_2)^2]_{\times 2}, \\[2mm]
	 [(m_1 \pm m_3 \pm m_4)^2]_{\times 2}, \ [(m_2 \pm m_3 \pm m_4)^2]_{\times 2} \\[2mm]
	\end{array}$ \\[2mm]
	$m^2_{scal}$ & $[0]_{\times 18}$,\ $\left[(m_1\pm m_2)^2\right]_{\times 2}$,\ $\left[(m_3\pm m_4)^2\right]_{\times 2}$,\ $\left[(m_1 \pm m_2 \pm m_3 \pm m_4)^2\right]_{\times2}$\\[2mm]
	\hline
\end{tabular}
\end{center}
\caption{Masses for the CSS vacuum M1.}
\end{table}

\begin{table}[H]
	\begin{center}
\rowcolors{1}{white}{gray!15}
\begin{tabular}{|c|c|}
	\hline
	$m^2_{3/2}$ & $[0]_2, \ [m_1^2]_2, \ [m_2^2]_2, \ [m_3^2]_2 $\\[2mm]
	$m^2_{vec}$ & $[0]_{3}, \ [(m_1\pm m_2)^2]_2, \ [(m_1\pm m_3)^2]_2, \ [(m_2\pm m_3)^2]_2,$\\[2mm]
	$m^2_{tens}$ & $ [m_1^2]_4, \ [m_2^2]_4, \ [m_3^2]_4,$ \\[2mm]
	$m^2_{1/2}$	&  $\begin{array}{c}
	[0]_{10},\ [m_1^2]_{2}, \ [m_2^2]_{2},\ [m_3^2]_{2}, \ [(m_1\pm m_2)^2]_{4}, \ [(m_1\pm m_3)^2]_4, \\[2mm]
	 \ [(m_2\pm m_3)^2]_{4}, \ [(m_1 \pm m_2 \pm m_3)^2]_2
	\\[2mm]
	\end{array}$ \\[2mm]
	$m^2_{scal}$ & $[0]_{14}, \ [m_1^2]_4, \ [m_2^2]_4,  \ [m_3^2]_4,\ [(m_1\pm m_2\pm m_3)^2]_4, $\\[2mm]
	\hline
\end{tabular}
\end{center}
\caption{Masses for the Minkowski vacuum M2.}
\end{table}

\begin{table}[H]
	\begin{center}
\rowcolors{1}{white}{gray!15}
\begin{tabular}{|c|c|}
	\hline
	$m^2_{3/2}$ & $[0]_{\times 4},  \ [m_1^2]_{\times 2}, \ [m_2^2]_{\times 2}$\\[2mm]
	$m^2_{vec}$ & $[0]_{\times 4},   \ [m_1^2]_{\times 4}, \ [m_2^2]_{\times 4}, \ [(m_1\pm m_2)^2]_{\times 2}$\\[2mm]
	$m^2_{tens}$ & $[0]_{\times 3},  \ [m_1^2]_{\times 4}, \ [m_2^2]_{\times 4}$\\[2mm]
	$m^2_{1/2}$	& $[0]_{\times 12},  \ [m_1^2]_{\times 10},\ [m_2^2]_{\times 10},\ [(m_1\pm m_2)^2]_{\times 8}$ \\[2mm]
	$m^2_{scal}$ & $[0]_{\times 18}, \ [m_1^2]_{\times 4}, \ [m_2^2]_{\times 4}, \ [(m_1\pm m_2)^2]_8$\\[2mm]
	\hline
\end{tabular}
\end{center}
\caption{Masses for the Minkowski vacuum M3.}
\end{table}

\begin{table}[H]
	\begin{center}
\rowcolors{1}{white}{gray!15}
\begin{tabular}{|c|c|}
	\hline
	$m^2_{3/2}$ &  $[m^2]_{\times 4}, \ [3m^2]_{\times 4}$ \\[2mm]
	$m^2_{vec}$ &  $[0]_{\times 4}, \ [4m^2]_{\times 10},  \ [8m^2]_{\times 3}$\\[2mm]
	$m^2_{tens}$ & $[0]_{\times 2}, \ [4 m^2]_{\times 8}$\\[2mm]
	$m^2_{1/2}$	&  $[0]_{\times 8},\ [m^2]_{\times 8},\ [3 m^2]_{\times 12}, \ [7 m^2]_{\times 8}, \ [9 m^2]_{\times 12}$ \\[2mm]
	$m^2_{scal}$ & $[0]_{\times 20},\ [4m^2]_{\times 10},\  [8m^2]_{\times 6}, \ [12m^2]_{\times 6}$\\[2mm]
	\hline
\end{tabular}
\end{center}
\caption{Masses for the Minkowski vacuum M4.}
\end{table}

\begin{table}[H]
	\begin{center}
\rowcolors{1}{white}{gray!15}
\begin{tabular}{|c|c|}
	\hline
	$m^2_{3/2}$ &  $[0]_{\times 4}, \ [m_1^2]_{\times 2}, \ [m_2^2]_{\times 2}$ \\[2mm]
	$m^2_{vec}$ &  $[0]_{\times 3},\ [m_1^2]_{\times 4}, \ [m_2^2]_{\times 4}, \ [m_1^2+m_2^2\pm m_3^2]_{\times 2}$\\[2mm]
	$m^2_{tens}$ & $[0]_{\times 4},\ [m_1^2]_{\times 4}, \ [m_2^2]_{\times 4}$\\[2mm]
	$m^2_{1/2}$	&  $[0]_{\times 12},\ [m_1^2]_{\times 10}, \ [m_2^2]_{\times 10}, \ [m_1^2+m_2^2\pm m_3^2]_{\times 8}$ \\[2mm]
	$m^2_{scal}$ & $[0]_{\times 18},\ [m_1^2]_{\times 4}, \ [m_2^2]_{\times 4}, \ [m_1^2+m_2^2\pm m_3^2]_{\times 8}$\\[2mm]
	\hline
\end{tabular}
\end{center}
\caption{Masses for the Minkowski vacuum M5.}
\end{table}

The other interesting fact that emerges from the spectra is that also in 5 dimensions, like in 4, Minkowski vacua have moduli.
In fact, once we remove the scalars that are eaten by the massive vectors in the usual Higgs mechanism, we see that the vacuum (M2) has two additional massless fields, the vacuum (M3) has 6 additional moduli, the vacuum (M4) 7 and the vacuum (M5) again 6.
Like in the 4-dimensional case \cite{Catino:2013ppa}, it may be worth investigating if  these gaugings can be connnected to each other by infinite distance limits along their moduli spaces.
Quite possibly, the most general such limits may also generate novel gaugings with new Minkowski vacua and residual symmetries other than U(2).



%

%
\bigskip
\section*{Acknowledgments}

\noindent We would like to thank T.~Fischbacher, C.~Krishnan and especially M.~Trigiante for useful correspondence.
This work is supported in part by the MIUR-PRIN contract 2017CC72MK003 \emph{``Supersymmetry breaking with fields, strings and branes''}.
This project has received funding from the European Union’s Horizon 2020 research and innovation programme under the Marie Skłodowska-Curie grant agreement No 842991.
%


\appendix

\section{Fermion shifts at the vacuum} 
\label{sec:fermion_shifts_at_the_vacuum}

In this appendix we provide an instance of the value of the fermion shifts generating the vacua of table 7.
For all examples we have chosen a basis where either 
\begin{equation}\label{W1}
		\Omega = {\mathbb 1}_4 \otimes i \, \sigma_2
\end{equation}
or
\begin{equation}\label{W2}
		\Omega = i \, \sigma_2 \otimes {\mathbb 1}_4.
\end{equation}

\textbf{A1}. In the basis with $\Omega$ as in (\ref{W1}), the maximal AdS supersymmetric vacuum is easily obtained by setting
\begin{equation}
		A_{16} = A_{38} = -A_{25} = - A_{47} = g, \quad A_{i,jkl} = 0.
\end{equation}

\bigskip

\textbf{A2}. In the basis with $\Omega$ as in (\ref{W1}), the SO(5) non-supersymmetric AdS vacuum follows from choosing
\begin{equation}
		A_{16} = A_{38} = -A_{25} = - A_{47} = g, 
\end{equation}
and
\begin{eqnarray}
		&& A_{1162} = A_{2251} = A_{3384} = A_{4473}= A_{5562} = A_{6651} = A_{7784} = A_{8873} = \frac{g}{4},  \\[2mm]
		&& A_{1238} = A_{1274} = A_{2183} = A_{2147} = A_{3164} = A_{3245} = A_{4136}= A_{4325} \nonumber \\[2mm]
		&&=A_{5368} = A_{5647} = A_{6385} = A_{6457} = A_{7861} = A_{7258} = A_{8167} = A_{8275} = \frac{3}{16}g, \\[2mm]
		&& A_{1364} = A_{2345} = A_{3182} = A_{4127} = A_{5278} = A_{6718} = A_{7456} = A_{8365} =  \frac{g}{16},  \\[2mm]
		&& A_{1678} = A_{2758} = A_{3568} = A_{4576} = A_{5243} = A_{6134}=A_{7214} = A_{8123} = \frac{5}{16}g. 
\end{eqnarray} 

\bigskip

\textbf{A3}. In the basis with $\Omega$ as in (\ref{W2}), the SU(3) invariant AdS vacuum follows from
\begin{equation}
		A_{15} = A_{26} = A_{37} = \frac79\,i\, m_1, \quad A_{48} =-i\, m_1,
\end{equation}
and
\begin{eqnarray}
		&& A_{1256}=A_{1357}=A_{2165}=A_{2367}=A_{3517}=A_{3276} \nonumber \\[2mm]
		&& =A_{5162}=A_{5317}=A_{6125}=A_{6273}=A_{7135}=A_{7236}=\frac{i}{9}\,g,\\[2mm]
		&& A_{1548}=A_{2648}=A_{3748}=A_{5148}=A_{6248}=A_{7348}=\frac{2}{9}\,i\,g,\\[2mm]
		&& A_{1234}=A_{2314}=A_{3124} = A_{5678}=A_{6587}=A_{7568} = \frac{1}{2\sqrt3}\,g,\\[2mm]
		&& A_{4123}=A_{8567}=\frac{\sqrt3}{2}\,g.
\end{eqnarray} 

\bigskip

\textbf{A4}. In the basis with $\Omega$ as in (\ref{W1}), the $N=2$ AdS vacuum with U(2) residual symmetry follows from
\begin{equation}
		A_{14} = -A_{23} = \frac{7}{12}\, g, \quad A_{56} = -i\frac{g}{2}, \quad A_{78} =i\,\frac23\,g,
\end{equation}
and
\begin{eqnarray}
		&& A_{1124} = A_{2213} = A_{3324} =A_{4413} =-\frac{5}{24}\,g,\\[2mm]
		&& A_{1275} = A_{1268} = A_{2157} =A_{2186} =A_{3457} =A_{3486} =A_{4375} =A_{4368} =\frac{i}{8}\,g,\\[2mm]
		&& A_{1475} = A_{1486} =A_{2357} =A_{2368} =A_{3275} =A_{3286} =A_{4157} =A_{4168} =\frac{g}{8},\\[2mm]
		&& A_{1456} = A_{2365} =A_{3265} =A_{4156} =\frac{g}{24},\\[2mm]
		&& A_{1478} = A_{2387} =A_{3287} =A_{4178} =\frac{g}{6}, \\[2mm]
		&& A_{7568} = A_{8567} = -i\,\frac{g}{6}, \\[2mm]
		&& A_{7152} = A_{7345} =A_{8126} =A_{8436} =i\,\frac{g}{4}, \\[2mm]
		&& A_{7154} = A_{7235} =A_{8416} =A_{8236} =\frac{g}{4}, \\[2mm]
		&& A_{7128} = A_{7348} =A_{8127} =A_{8347} =i\,\frac{g}{12}. 
\end{eqnarray} 

\bigskip

\textbf{A5}. In the basis with $\Omega$ as in (\ref{W1}), the $N=0$ AdS vacuum with SU(2) $\times$ U(1)$^2$ residual symmetry follows from
\begin{equation}
		A_{23} = -A_{14} = \frac{1}{3}\sqrt{\frac25}\, g, \quad A_{56} = -i\frac{g}{5}, \quad A_{77}=-A_{88} =i\,\frac{g}{5},
\end{equation}
and
\begin{eqnarray}
		&& A_{1124}=A_{2213}=A_{3324}=A_{4413}=\frac{g}{3\sqrt{10}},\\[2mm]
		&& A_{1456}=A_{2365}=A_{3287}=A_{4178}=\frac{1}{12}\sqrt{1-\frac25\sqrt6}\,g,\\[2mm]
		&& A_{1487}=A_{2378}=A_{3256}=A_{4165}=\frac{1}{12}\sqrt{1+\frac25\sqrt6}\,g,\\[2mm]
		&& A_{5164}=A_{5236}=A_{6145}=A_{6253}=A_{7148}=A_{7283}=A_{8174}=A_{8237}=\frac{g}{4\sqrt{15}},\\[2mm]
		&& A_{5621}=A_{6521}=A_{7734}=A_{8843}=i\,\frac{1}{60}\left(2 +\sqrt6\right)g, \\[2mm]
		&& A_{5634}=A_{6534}=A_{7721}=A_{8812}=i\,\frac{1}{60}\left(-2 +\sqrt6\right)g, \\[2mm]
		&& A_{5678}=A_{6578}=A_{7765}=A_{8856}=i\,\frac{g}{15}. 
\end{eqnarray} 

\bigskip

\textbf{M1}. The general CSS Minkowski vacuum in the basis with $\Omega$ as in (\ref{W1}), follows from 
\begin{equation}
		A_{11}=A_{22} = \frac{m_1}{3}, \ A_{33} = A_{44}  =  \frac{m_2}{3}, \ A_{55} = A_{66}  =  \frac{m_3}{3}, \ A_{77} = A_{88}  =  \frac{m_4}{3}, \ 
\end{equation}
and
\begin{eqnarray}
		&& A_{1134} = A_{2234} = -\frac{m_1}{3}, \\[2mm]
		&& A_{1156} = A_{1178} =A_{2256}= A_{2278} = \frac{m_1}{6}, \\[2mm]
		&& A_{3312} = A_{4412} = -\frac{m_2}{3}, \\[2mm]
		&& A_{3356} = A_{3378} =A_{4456}= A_{4478} = \frac{m_2}{6}, \\[2mm]
		&& A_{5578} = A_{6678} = -\frac{m_3}{3}, \\[2mm]
		&& A_{5512} = A_{5534} =A_{6612}= A_{6634} = \frac{m_3}{6}, \\[2mm]
		&& A_{7756} = A_{8856} = -\frac{m_4}{3}, \\[2mm]
		&& A_{7712} = A_{7734} =A_{8812}= A_{8834} = \frac{m_4}{6}.
\end{eqnarray} 
Obviously the vacua that appear in the context of our analysis have some of the masses either set to zero or proportional to each other, in order to respect the correct U(2) residual symmetry, but they are always subcases of the one presented here.

\textbf{M2}. The Minkowski vacuum from the SU(3,1) gauging appears in the basis with $\Omega$ as in (\ref{W1}) by choosing
\begin{equation}
		A_{34}= i \frac{m_1}{3}, \ A_{56} =  i \frac{m_2}{3}, \ A_{77} = A_{88}  =  \frac{m_3}{3}
\end{equation}
and
\begin{eqnarray}
		&& A_{3124} = A_{4123} = -i\,\frac{m_1}{3}, \\[2mm]
		&& A_{3456} = A_{3478} =A_{4356}= A_{4378} = i\,\frac{m_1}{6}, \\[2mm]
		&& A_{5126} = A_{6125} = -i\,\frac{m_2}{3}, \\[2mm]
		&& A_{5346} = A_{5678} =A_{6345}= A_{6578} = i\,\frac{m_2}{6}, \\[2mm]
		&& A_{7712} = A_{8812} = -\frac{m_3}{3}, \\[2mm]
		&& A_{7734} = A_{7756} =A_{8834}= A_{8856} = \frac{m_3}{6}.
\end{eqnarray} 

\textbf{M3}. The first new Minkowski vacuum we found appears in the basis with $\Omega$ as in (\ref{W1}) by choosing
\begin{equation}
		A_{56} =  -i \frac{m_2}{3}, \ A_{77} = A_{88}  =  \frac{m_1}{3}
\end{equation}
and
\begin{eqnarray}
		&& A_{5346} = A_{6345} =i\,\frac{m_2}{3}, \\[2mm]
		&& A_{5126} = A_{5678} = A_{6125}= A_{6578} = -i\,\frac{m_2}{6}, \\[2mm]
		&& A_{7734} = A_{8834} = -\frac{m_1}{3}, \\[2mm]
		&& A_{7712} = A_{7756} =A_{8812}= A_{8856} = \frac{m_1}{6}.
\end{eqnarray} 

\textbf{M4}. The new non-supersymmetric Minkowski vacuum appears in the basis with $\Omega$ as in (\ref{W1}) by choosing
\begin{equation}
		A_{23}= - A_{14} = \frac{m_1}{3}, \ A_{58} = -A_{67}  =  \frac{m_1}{\sqrt3}
\end{equation}
and
\begin{eqnarray}
		&& A_{1142} = A_{1478} = A_{2231} = A_{2387} = A_{3342} = A_{3287} = A_{4431} = A_{4178} = \frac{m_1}{6}, \\[2mm]
		&& A_{1485} = A_{1467} = A_{2358} = A_{2376} = A_{3258} = A_{3276} = A_{4185} = A_{4167} = \frac{m_1}{3\sqrt2}, \\[2mm]
		&& A_{5182} = A_{5384} = A_{6127} = A_{6347} = A_{7126} = A_{7346} = A_{8215} =A_{8435} =  \frac{m_1}{4\sqrt3}, \\[2mm]
		&& A_{5146} = A_{5236} = A_{6145} = A_{6235} = A_{7148} = A_{7238} = A_{8147} = A_{8237} = \frac{m_1}{4}, \\[2mm]
		&& A_{5568} = A_{6657} = A_{7768} = A_{8857} = \frac{m_1}{2\sqrt3}\\[2mm]
		&& A_{5542} = A_{6631} = A_{7742} = A_{8831} = \frac{m_1}{2}.
\end{eqnarray}

\textbf{M5}. The new $N=4$ Minkowski vacuum appears in the basis with $\Omega$ as in (\ref{W1}) by choosing
\begin{equation}
		A_{56}=-i \,\frac{m_2}{3}, \ A_{77} = A_{88}  =  \frac{m_1}{3}
\end{equation}
and
\begin{eqnarray}
		&& A_{5126} = A_{6125} = i\,\frac{m_2}{3}, \\[2mm]
		&& A_{5346} = A_{5678} = A_{6345} = A_{6578} = -i\, \frac{m_2}{6},\\[2mm]
		&& A_{7712} = A_{8812} = -\frac{m_1}{12} - \frac{m_3^2}{8 m_2}, \\[2mm]
		&& A_{7734} = A_{8834} = -\frac{m_1}{12} + \frac{m_3^2}{8 m_2}, \\[2mm]
		&& A_{7714} = A_{7732} = A_{8814} = A_{8832} = -\frac18\,\sqrt{4m_1^2-\frac{m_3^4}{m_2^2}}, \\[2mm]
		&& A_{7756} = A_{8856} = \frac{m_1}{6}. 
\end{eqnarray} 

\textbf{D1}. The de Sitter vacuum associated to the SO(3,3) gauging appears in the basis with $\Omega$ as in (\ref{W2}) by choosing
\begin{equation}
		A_{ij} = 0
\end{equation}
and
\begin{eqnarray}
		&& A_{1278} = A_{1386} =A_{1476} =A_{2187} =A_{2358} =A_{2457} =A_{3168} =A_{3285} =m, \\[2mm]
		&& A_{3465} =A_{4167} =A_{4275} =A_{4356} =A_{5283} =A_{5274} =A_{5346} =A_{6138} = m, \\[2mm]
		&& A_{6147} =A_{6354} =A_{7182} =A_{7164} =A_{7245} =A_{8127} =A_{8163} =A_{8235} = m. 
\end{eqnarray} 

\textbf{D2}. The new de Sitter vacuum associated to the SU(3,1) gauging appears in the basis with $\Omega$ as in (\ref{W2}) by choosing
\begin{equation}
		A_{15} = A_{26} = A_{37} = 3\,i\,g, \quad A_{48} = -i\, g,
\end{equation}
and
\begin{eqnarray}
		&& A_{1265} = A_{1375} = A_{2156} = A_{2376} = A_{3157} = A_{3267} \nonumber \\[2mm]
		&&= A_{5126} = A_{5137} = A_{6152} = A_{6237} = A_{7153} = A_{7263} = i\,g, \\[2mm]
		&& A_{1458} = A_{2468} = A_{3478} = A_{5184} = A_{6284} = A_{7384} = 2i\,g,\\[2mm]
		&& A_{1287} = A_{1368} = A_{2178} = A_{2385} = A_{3186} = A_{3258} = A_{4167} = A_{4275} = A_{4356}  \\[2mm]
		&&= A_{5247} = A_{5364} = A_{6174} = A_{6345} = A_{7146} = A_{7254} = A_{8127} = A_{8163} = A_{8235} = \frac{\sqrt{3}}{2}\,g, \nonumber\\[2mm] 
		&& A_{1467} = A_{2475} = A_{3456} = A_{5238} = A_{6183} = A_{7128} = \frac{3\sqrt{3}}{2}\,g. 
\end{eqnarray}


%

%
%
%

\end{document}